\documentstyle[aps,prl,twocolumn,epsf,floats]{revtex}
	
\newcommand{\beq}{\begin{equation}}
\newcommand{\eeq}{\end{equation}}
\newcommand{\bea}{\begin{eqnarray}}
\newcommand{\eea}{\end{eqnarray}}

\newcommand{\etal}{{\it et al.},}
\newcommand{\nn}{\nonumber}

\begin{document}

\title{\bf \Large
The spatial profile of the matter distribution in a
dilute atomic-molecular Fermi cloud
}
\vspace{0.75cm}
\author{ Aurel Bulgac  }
\vspace{0.50cm}

\address{Department of Physics, University of
Washington, Seattle, WA 98195--1560, USA}

\maketitle

\begin{abstract}

A dilute atomic-molecular fermion in a trap acquires a rather complex
layered spatial structure. At very low temperatures, typically at the
center of the trap, a superfluid bosonic core is formed out of dimers,
weakly bound pairs of fermionic atoms. Surrounding this core is a
mantle, formed by a mixture of superfluid bosonic dimers and
superfluid fermionic atoms. The outermost layer consists of only
fermionic atoms, forming mostly a normal Fermi gas.

\end{abstract}

\draft


%
%
%
%
%

Jin's group \cite{jin} has demonstrated that weakly bound dimers of
fermionic atoms are easily created in a cloud of cold Fermi atoms, by
ramping up and down the magnetic field near a Feshbach resonance.
Several other groups have been able to quickly confirm these results,
by using similar experimental techniques
\cite{grimm,hulet,salomon}. In all these experiments a relatively cold
Fermi gas of atoms, interacting with a negative scattering length, was
driven past the Feshbach resonance into a mixture of fermionic atoms
and bosonic dimers, all of them interacting with positive scattering
lengths \cite{abpfb_f}. The ratio of atoms to molecules in these
experiments could be varied essentially at will, by tuning the ramping
rate of the ambient magnetic field.  It was also explicitly
demonstrated that this mixed system of fermionic atoms and bosonic
dimers has a rather long lifetime \cite{grimm,regal} and that the rate
of various recombination processes is rather low, an effect
qualitatively suggested in Ref. \cite{abpfb_f}. In a series of related
experiments it was shown that the weakly bound dimers form a
Bose-Einstein condensate \cite{mbec}.

The natural question arises: "What is the nature of this new quantum
object created in these experiments?" As it was shown in
Ref. \cite{abpfb_f}, in the regime where the scattering length is
large, but at the same time the various average inter-particle
separations are even larger, it is possible to perform a rather
straightforward theoretical analysis of its properties. If the system
could be regarded as locally homogeneous, then by changing the
scattering length from negative to positive values part of the system
undergoes a BCS $\rightarrow$ BEC crossover
\cite{leggett,nozieres,randeria,mohit}.  At zero temperature, when the
scattering length is negative, one would expect that the atomic Fermi
system would be in a BCS state at sufficiently low temperatures.  By
changing adiabatically the scattering length to positive values, the
size of the Cooper pairs would gradually decrease, until dimers, with
sizes smaller than the average inter-particle separation, are
eventually formed and the entire system turns into a pure BEC. At the
finite temperatures achieved so far in atomic Fermi clouds, typically
only a fraction of the atoms are converted into dimers and one ends up
with a mixed system of fermions and bosons.

In all these experiments \cite{jin,grimm,hulet,salomon} the estimated
final temperatures were well below the critical BEC temperature of the
Bose component.  One could have naively concluded that a novel mixture
of a Bose superfluid mixed with a normal Fermi gas was thus realized.
One reason to believe this would be the fact that the atomic
scattering length is positive, which corresponds to an effective
repulsion at these energies. Therefore the Fermi subsystem could be
expected to be a normal Fermi system.  The other reason would be the
fact that the dimers were expected to repel each other
\cite{abpfb_f,mohit,pieri,petrov} and thus they can form a BEC. This aspect
was subsequently confirmed experimentally \cite{mbec}. A third reason
would be the fact that atoms and dimers repel under these
conditions too \cite{abpfb_f,petrov}.

Nevertheless, a somewhat subtle quantum phenomenon could lead to
rather dramatic changes of the properties of the fermionic
component. Namely, even though two fermions effectively repel each
other (since the scattering length is positive), there is a boson
induced interaction, which is attractive and relatively long ranged. A
fermion traveling in a Bose superfluid (fermions always move) will
excite a Bogoliubov sound mode of the Bose subsystem, which in its
turn will propagate to rather large distances, where it will "shake"
another fermion. This type of interaction between fermions is
reminiscent of the phonon mediated electron-electron interaction in
metals, which is responsible for the superconductivity. Technically,
this phonon exchange between two fermions is a second order
perturbation theory contribution to the ground state energy of the
system, and as such is always attractive. As a result, if the
temperatures are low enough and the induced attraction strong enough,
a homogeneous mixture of such bosons and fermions could turn into a
mixtures of a Bose superfluid in interaction with a Fermi superfluid.
If the system is dilute and if all the scattering lengths are known
\cite{abpfb_f} and the energy of such a system is easily
calculated. Here, I shall primarily be concerned with the spatial
profile and spatial structure of such a mixed atom-dimer system at low
temperatures. Since the induced pairing correlations in the Fermi
subsystem are in the weak coupling regime \cite{abpfb_f}, the
presence or absence of the pairing correlations plays essentially no
role in determining the spatial distribution of such an atomic cloud.

The properties of these novel Bose-Fermi mixtures of quantum fluids
could be affected by the shape of the trapping potential in a
non-trivial manner.  As it will be shown here, the trapping potential
has a rather strong influence on the detailed spatial matter density
profile and spatial composition of these new objects.  Even though one
can create either a pure Bose system or a pure Fermi system, however,
such objects are likely an exception rather than the rule.  Most of
the time one would create an object, which at the center will be a
pure Bose superfluid, followed by a layer of mixed Bose and Fermi
superfluids, with an outer layer being a normal Fermi fluid.  At the
interface between the normal and the superfluid Fermi systems, another
pure quantum phenomenon, known as Andreev reflection, leads to the
formation of superfluid correlations in the outer normal Fermi fluid.
These pairing correlations will be present in a layer of width
comparable with the so called coherence length \cite{andreev}.  When a
fermion, impinging on a normal-superfluid interface from the normal
part, undergoes an Andreev reflection, it is reflected essentially
backward as a hole. This process leads to the appearance in a
relatively narrow region of the normal Fermi fluid of pairing
correlations, even though the atom-atom interaction is repulsive.
Again, the presence of such pairing correlations will have an
exceedingly small influence on the spatial distribution of the atoms 
\cite{abyy}.

In Ref. \cite{abpfb_f} it was shown that the energy density of a mixed
fermion-dimer system in a trap is given by the following expression:
\bea
& &  {\mathcal{E}}= \frac{3}{5}\frac{\hbar^2k_F^2}{2 m}n_f
+\frac{1}{2}U_{ff}n_f^2  \nn \\
& &  +  U_{fb}n_fn_b
+\frac{1}{2}U_{bb}n_b^2 +\varepsilon_2 n_b + Vn_f +2Vn_b , \label{eq:edf}
\eea 
where $V$ is the trap potential, $n_b$ is the dimer number density,
$n_f=k_F^3/3\pi^2$ is the total number density of both fermion
species, assumed here to be present in equal amounts, $k_F$ is the
Fermi wave vector and where
\bea
& & \varepsilon_2=-\frac{\hbar^2}{ma^2}, 
\quad  U_{ff}=2     \; \frac{\pi \hbar^2 a}{m}>0, \\
& & U_{fb}=3.537 \; \frac{\pi \hbar^2 a}{m}>0,  
\quad  U_{bb}=1.2   \; \frac{\pi \hbar^2 a}{m}>0
\eea 
and $a$ is the atom-atom scattering length.  The trapping potential
experienced by a dimer has twice the value of the corresponding
potential experienced by an atom at the same position. The position
dependence of the number densities and trap potentials are not shown
explicitly.  Since various recombination rates are sufficiently small
\cite{grimm,abpfb_f,regal,petrov} one can consider that the number of
fermions and bosons are conserved separately and one can account for
the corresponding slow changes later-on when and if needed. As it was
mentioned a few times above and it was shown in Ref. \cite{abpfb_f},
the fermions, in the presence of superfluid bosons, become superfluid
at sufficiently low temperatures. The gain in the energy density
arising from the appearance of superfluid correlations could in
principle be included in the above energy density.  However,
this gain is very  modest and is not expected to lead to any
major changes of the spatial profile of this system. This gain in
energy is called condensation energy and per particle is
given by $3\Delta^2/8\varepsilon_F\ll \varepsilon_F$,
where $\varepsilon_F$ is the Fermi energy.

In the case of a large number of atoms one can find the atom and dimer
number densities with excellent accuracy in the Thomas-Fermi
approximation \cite{collective,grasso}, namely form the equations:
\bea
& & \frac{\hbar^2k_F^2}{2 m}
+\left ( U_{ff}-\frac{U_{fb}^2}{U_{bb}}\right ) n_f \nn \\
& &  = \mu_f -\frac{ U_{fb} }{ U_{bb} }\mu_b -
\left ( 1-\frac{2U_{fb}}{U_{bb}} \right ) V ,   \label{eq:fermi}\\
& & n_b = \frac{\mu_b-2V -U_{fb}n_f}{U_{bb}}. \label{eq:bose} 
\eea 
I have absorbed in the definition of the chemical potential $\mu_b$
the energy of the dimer $\varepsilon_2$. These two equations have been
obtained by assuming that both $n_{f,b}\neq 0$.  If either one of the
number densities is vanishing, then one has to solve instead one of
the following two equations:
\bea
& & \frac{\hbar^2k_F^2}{2 m} +U_{ff}n_f = \mu_f -V, 
\quad {\mathrm if} \quad n_f > 0 
\quad {\mathrm and} \quad n_b\equiv  0\label{eq:ff} \\
& & n_b = \frac{\mu_b-2V}{U_{bb}},  
\quad {\mathrm if} \quad n_b > 0 
\quad {\mathrm and} \quad n_f\equiv  0
 \label{eq:bb} 
\eea 
As usual in the Thomas-Fermi approximation, in the region of space
where densities are exponentially small (classically forbidden
regions), the corresponding densities are set equal to zero.  In
principle one can rather straightforwardly improve on this
approximation, but changes in distributions are quantitatively minor.
Since $1-2U_{fb}/U_{bb}\approx -5$, see Eq. (\ref{eq:fermi}), the
effective trap potential experienced by fermions changes its character
in the presence of dimers and the fermions are pushed outward.  In the
spatial region where both $n_f\neq 0$ and $n_b\neq 0$ the fermion
number density increases outwardly.  By analyzing various
possibilities one is quickly convinced that besides the rather
rare cases, when there are either only fermions or only bosons, a finite
cloud consisting of both Fermi atoms and Bose dimers has fermions only
in the outside layer.

I shall describe now the profiles of the atom and dimer number
densities, by constructing them from point to point. I
assume that the trap potential is increasing monotonically from its
"center" outward. The "center" of the trap is defined as the point
where the trapping potential has a minimum.  The shape of the potential
could otherwise be arbitrary. I shall fix the origin of the coordinate
system in the "center" of the trap and  construct the number
densities along an arbitrary straight line. Let me denote by
$\bbox{R}_3$ the farthest point from origin along this line where the
atom density vanishes, that is where $n_f(\bbox{R}_3)=0$. At this
point
\beq
\mu_f = V(\bbox{R}_3), \label{eq:muf}
\eeq
and one can thus fix the value of the atom chemical
potential. According to Eq. (\ref{eq:muf}) the atom number density
vanishes along an equipotential surface.  Starting from the point
$\bbox{R}_3$ one moves inward and point by point solves
Eq. (\ref{eq:ff}) and thus determine the atom number density along
this chosen line. Now I shall assume that there are dimers present in
the system as well, and that their number density vanishes along the
same line at the farthest point from origin $\bbox{R}_2$ ($R_2<R_3$).
According to Eq.  (\ref{eq:bose}) at this point the following
condition is fulfilled
\beq
\mu_b = 2V(\bbox{R}_2)+U_{fb} n_f(\bbox{R}_2), \label{eq:mub}
\eeq
which thus determines the value of the dimer chemical potential.
Again, the locus of points where $n_b(\bbox{R}_2)=0$ is another
equipotential surface of the trapping potential.  From this point
inward, one has now to solve the equations (\ref{eq:fermi}) and
(\ref{eq:bose}).  One can then find a point $\bbox{R}_1$ ($R_1<R_2$)
along this chosen line, where $n_f(\bbox{R}_1)=0$. Such a point might
or might not exist, and its presence is determined by the particular
values of the chemical potentials $\mu_{f,b}$, and therefore by the
particular total number of atoms and dimers.  Naturally, if the number
of dimers is sufficiently large, such a point $\bbox{R}_1$, where
$n_f(\bbox{R}_1)=0$, exists.  If such a point $\bbox{R}_1$ exists,
from that point on inward the atom number density vanishes identically
and the dimer number density is determined from Eq. (\ref{eq:bb})
alone. Moreover, the dimer number density continues to increase
further on from the point $\bbox{R}_1$ until it reaches the origin.
The equipotential surface, which separates the bosons only region at
the center of the trap, from the region with both fermions and bosons
can be determined from the equation:
\beq
U_{fb}\mu_b-U_{bb}\mu_f = (2U_{fb}-U_{bb})V(\bbox{R}_1). \label{eq:mufb}
\eeq
One could have generated the same atom and dimer density profiles by
starting from the origin instead.  In that case one chooses a
particular value for the dimer chemical potential $\mu_b$. One has
however at that point to decide whether at the origin one has atoms as
well or not. Depending on the presence of atoms at the origin, one
proceeds in the reverse order and generates both atom and dimer number
densities.

Once the number densities have been determined along one chosen line,
one simply extends the values of the number densities along
equipotential surfaces in the entire space. This overall onion
structure is very reminiscent of the structures one has inside the
Earth or neutron stars, with a core of one nature, followed by a
mantle, which has a structure of a different nature, and eventually by
a crust with a different structure as well.  In the present case
however all these three substructures are fluid. The extension of the
present analysis to finite temperatures is straightforward, and one
has to consider the free energy density, instead of the energy
density, by adding the corresponding contributions arising from the
fermionic and bosonic entropies as well as allowing for the bosons to
have a non-condensed fraction. I shall limit the analysis here
to the zero temperature case only.  It is expected that with
increasing temperature, first the fermionic superfluidity of the
mantle is lost and subsequently the superfluidity of the bosonic core.

Some rather simple relations between these various radii can be
obtained by taking into account the fact that the system is dilute,
thus $n_fa^3 \ll 1$ and $n_fba^3 \ll 1$. One then easily obtains from
Eqs. (\ref{eq:muf} - \ref{eq:mufb}) that the three radii discussed above
in a harmonic trap are related (with a 2\% error)
\beq
6 R_2^2 =   R^2_3 + 5 R^2_1.
\eeq
Even though terms like $U_{fb}n_f$ have been dropped when this
relation was obtained, the interaction is responsible for the specific
numbers appearing here, as well as for the existence of this relation.
In particular, this relations shows that the inner core disappears as
soon as $R_2 < R_3/\sqrt{6}$.  The atom and dimer number density
distribution are then given in a spherical trap by, see Fig. 1,
\bea
n_f& =& \frac{1}{3\pi^2}\left (\frac{m\omega}{\hbar}\right )^3
( 0.2\; R_3^2 - 1.2 \; R_2^2 + r^2)^{3/2}, \\
& & \quad \quad \quad  \quad \quad \quad
{\mathrm if} \quad R_1\leq r\leq R_2, \nn \\
n_f& =& \frac{1}{3\pi^2}\left (\frac{m\omega}{\hbar}\right )^3
( R_3^2 - r^2)^{3/2},   \quad 
{\mathrm if}  \quad R_2\leq r\leq R_3,  \\
n_b&=& \frac{1}{1.2\pi a} \left (\frac{m\omega}{\hbar}\right )^2
(R_2^2-r^2), \quad  {\mathrm if} \quad r\leq R_2.
\eea
The magnitude, but not the shape, of the dimer distribution depends on
the particular value of the scattering length. And since it has a very
simple shape it is not shown in this figure.  The specific numbers
here are universal in the regime of large positive scattering lengths,
when $a \gg r_0=(C_6m/\hbar^2)^{1/4}$. Here $\omega $ is the trap
frequency and $C_6$ is the strength of the atom-atom van der Waals
interaction.  The atom distribution is thus essentially independent of
the particular value of the scattering length $a$, as long as this is
large and the system is dilute. One can easily derive analytical
formulas for the number of atoms and dimers as functions of the radii
$R_{1,2,3}$, even for anisotropic traps.  Once the entire system is in
the quantum degenerate regime the temperature dependence of the number
density profiles are rather weak \cite{collective}. The ratio of
dimers to atoms scales as a function of $\omega$, $a$ and cloud radii
as follows
\beq
\frac{N_{dimer}}{N_{atom}}\propto \frac{\hbar  R_2^5 }{m\omega a R^6_3}.
\eeq
 Since the scattering length $a$ is the smallest length scale typically
 the number of dimers is larger than the number of atoms, but this is
 not true always.

\begin{figure}
\epsfxsize=7.0cm
\centerline{\epsffile{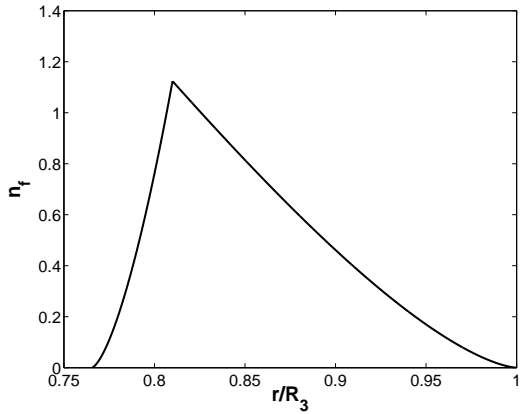}}
\caption{ \label{fig:fig1} 
The atom density profile of an atom-dimer mixture in the dilute limit,
when $R_2 = 0.81 \; R_3$ and where the number density was normalized
to one.  In this case about 16\% of the atoms are in the ``mantle.''
}
\end{figure}

Besides the rather intricate structure of this
new quantum object, one can naturally
expect that the spectrum of collective states is going to have a
nontrivial character, arguably different from the spectrum of BEC
systems studied so far \cite{collective,cwk}. As noted above, the core
and the mantle of this new quantum object are superfluid, the first as
a Bose system and the latter as mixed Bose-Fermi superfluids.  As the
most notable feature of a superfluid system is the possibility to
excite a superflow, it would be extremely interesting to study the
character of one or more vortex states. Unlike  Bose
\cite{cwk,vort_b} or Fermi \cite {vort_f} clouds, since the crust here
is a normal Fermi fluid, a vortex will have to be limited to either
the core and mantle or the mantle alone, and not extend into the
Fermi normal crust (apart from the proximity effects due to the
Andreev reflection mentioned above). It is likely that the Bose and
Fermi superfluids can sustain largely independent superflows and the
presence of a vortex in one component would not necessarily lead to
the appearance of a vortex in the other component as well. The
rotational properties of this object are clearly nontrivial, as one
can expect both rotational and irrotational flows at the same
time. The kinetics and thermalization of an object with such a complex
structure is going to provide a interesting challenge to unravel, both
experimentally and theoretically. However, unlike stars and planets,
we should be able to study these processes under a variety of
conditions and be able to finely tune various properties of these
"stars in a bottle."

In summary, it was shown that an atomic cloud consisting of fermionic
atoms in a trap, at sufficiently low temperatures and when the
atom-atom scattering length is positive and large, will have an
onion-like structure, with a superfluid bosonic core made of atomic
dimers, a mantle consisting of a mixture of superfluid bosonic dimers
and fermionic atoms, all surrounded by an outer layer consisting of
fermions alone, forming a normal Fermi gas.  The existence of this
particular structure is due to a large extent to the fact that various
couplings ``conspired'' to acquire the specific values quoted here,
see Eqs. (3 - 5).

 This work benefited from partial financial support under DOE contract
 DE--FG03--97ER41014.


\end{document}